# On the existence of pure, broadband toroidal sources in electrodynamics

Adrià Canós Valero[1*], Dmitrii Borovkov[2], Mikhail Sidorenko[1], Pavel Dergachev[3,4], Egor Gurvitz[1], Lei Gao[5], Vjaceslavs Bobrovs[2], Andrey Miroshnichenko[6] and Alexander S. Shalin[2,5,7,8,9*]

[1]ITMO University, St. Petersburg 197101, Russia
[2]Riga Technical University, Institute of Telecommunications, Riga 1048, Latvia
[3]National Research University Moscow Power Engineering Institute 14 Krasnokazarmennaya st., Moscow 111250, Russia
[4]National University of Science and Technology (MISIS)
4 Leninsky pr., Moscow 119049, Russia
[5] School of Optical and Electronic Information, Suzhou City University, Suzhou 215104, China
[6] School of Engineering and Information Technology University of New South Wales Canberra Campbell, ACT 2600, Australia
[7]Faculty of Physics, M. V. Lomonosov Moscow State University, 119991 Moscow, Russia
[8]Center for Photonics and 2D Materials, Moscow Institute of Physics and Technology, Dolgoprudny 141700, Russia
[9]Kotelnikov Institute of Radio Engineering and Electronics, Ulyanovsk 432000, Russia
*adria.canos5@gmail.com
*alexandesh@gmail.com

**Abstract**

Multipoles are paramount for describing electromagnetic fields in many areas of nanoscale optics, playing an essential role for the design of devices in plasmonics and all-dielectric nanophotonics. Challenging the traditional division into electric and magnetic moments, toroidal moments have been proposed as a physically distinct family of multipoles with significant contributions to the properties of matter. However, the apparent impossibility of separately measuring their response sheds doubt on their true physical significance. Here, we confirm the possibility of selectively exciting toroidal moments without any other multipole. We develop a set of general conditions that any current distribution must fulfill to be entirely described by toroidal moments and prove our results in an analytically solvable case. Our new theory allows us to design and verify experimentally for the first time an artificial structure supporting a pure broadband toroidal dipole response in the complete absence of the electric dipole and other 'ordinary' multipole contributions. In addition, we propose a structure capable of supporting a novel type of nonradiating source, a 'toroidal anapole,' originating from the destructive interference of the toroidal dipole with the unconventional electromagnetic sources known as mean square radii. The results in this work provide conclusive evidence on the independent excitation of toroidal moments in electrodynamics.

**Introduction**

Multipole expansions constitute a central framework in many areas of modern physics and underpin our understanding of several phenomena in electrodynamics, nuclear physics, and condensed matter[1–3]. In optics, multipole expansions are widely exploited to characterize the radiation properties of a finite source. Any multipolar formalism typically involves electric and magnetic dipoles and increasingly more complex combinations, such as quadrupoles, octupoles, and so forth[3,4]. It is nowadays well-known that the so-called 'Cartesian' multipole expansions, based on a Taylor series of the vector potential[5,6] or equivalently the internal current distribution[7], require the introduction of dynamic toroidal moments for a complete parametrization of any charge-current distribution, both classical and quantum[5]. The distinct constituents of this extended class of moments are generated by transverse current distributions associated with the curl of a magnetization vector, topologically similar to the poloidal currents flowing through toroidal surfaces[5,8,9].

Toroidal multipole moments have been shown to be important in several branches of physics where toroidal topology plays a role. The most recognizable member of the toroidal family is the toroidal dipole (TD), first identified [10] and observed in a static current distribution. In condensed matter systems, the static TD was shown to naturally give rise to a new type of long-range ferroic order violating both space and time inversion symmetries[1,10]. Similarly, in molecular physics, static toroidal moments were predicted to appear in the electronic structure of several molecules[2].

In modern electrodynamics, *dynamic* toroidal moments describe currents that vary in time. They are currently raising a renewed scientific and technological interest. In particular, they are important for describing light scattering from nanostructured matter in nanophotonics and meta-optics. By making use of artificially tailored media (metamaterials and their flat counterparts, metasurfaces), the ordinary electric dipole (ED) can be suitably designed to interfere destructively with the dynamic TD in the far field, leading to the formation of nontrivial, nonradiating sources, (dynamic anapoles[6,11]), with enhanced near fields. Anapoles have also been proposed as candidates for observing the dynamic Aharonov-Bohm effect [13]. Beyond the dynamic TD, higher order toroidal moments, also known as toroidal mean-square radii (MSR), have been recently proposed to realize novel types of nonradiating anapole states [6,13].

The first observation of a dynamic TD in a plasmonic metamaterial [14], triggered many investigations aiming to isolate it and understand its fundamental properties. Nowadays, structures featuring a certain degree of dynamic TD contribution have been designed for virtually all frequency bands of interest. The required current distribution of the source can therefore have many physical origins, e.g., localized plasmons [15], surface plasmon polaritons [16], photonic (Mie) resonances [17,18], or spoof surface plasmons [19]. In extended photonic metasurfaces, some eigenmodes with dynamic TD character correspond to symmetry-protected quasi-bound states in the continuum that couple weakly to the incident field, manifesting as a sharp Fano resonance in the transmission and reflection spectra [20]. The latter holds promising prospects for enhancing light-matter interactions and sensing applications in both plasmonic [21], all-dielectric [22–25], and hybrid plasmonic-dielectric platforms. Dynamic TD metadevices have been shown to outperform their elementary counterparts for a variety of tasks, such as polarization conversion [26], strong plasmon-exciton coupling [27], metasurface-enhanced waveguiding [28], or active optical switching [29].

However, the physical significance of the dynamic TD has been the subject of an active scientific debate[8,30,31]. The main source of controversy is centered around the fact that no measurement of the fields *radiated or scattered* by a dynamic TD can distinguish them from the ordinary ED (and can be related to the *observability* condition of dynamical systems). The statement applies to all measurements performed outside the smallest volume enclosing the source. This poses the question of whether there is any real need to define two separate quantities. More formally, as will be detailed later, the TD arises as the second order term in a Taylor series of the exact electric coefficients [30]. The split of such coefficient is questionable since it does not lead to additional observables outside the source region. The debate has also motivated interesting theoretical proposals on how the ordinary ED and the dynamic TD 'response' could be distinguished in an actual physical system in the far field [32]. Unfortunately, these ideas remain untested so far.

Currently, three significant drawbacks are preventing a more thorough investigation of dynamic toroidal moments. Firstly, due to the absence of clear, rigorous rules justifying the separation of the ordinary ED and the dynamic TD, it is impossible to confidently claim the observation of a 'pure' or 'ideal' dynamic TD. Secondly, in most structures investigated until now, the dynamic TD response is often masked by the contributions of other electric and magnetic multipoles[15,33–39]. Thirdly, while some works have achieved a significant background-to-noise ratio of the dynamic TD with respect to the other multipoles [14,17,19,20,40], they rely on isolated resonances, limiting their operation to a reduced frequency range. Moreover, even in the most well-known examples, the dynamic TD is accompanied by an ordinary ED with a similar lineshape as the TD but suppressed amplitude [14,17,19,20,40], hinting at the fact that the resonances are not 'pure,' and shedding doubt whether the dynamic TD can be truly separated from the ordinary ED.

Here, we aim to demonstrate that the excitation of an ideal dynamic TD is indeed possible and relies on physically distinct constraints with respect to the ordinary ED. To do so, we first determine the exact conditions that a current distribution of *arbitrary spatial extension* must satisfy to excite a pure dynamic TD in the complete absence of the two most common caveats, i.e., ordinary ED and parasitic magnetic moments. Besides being divergence-less, we show that the curl of such currents must not have a radial component and importantly cannot induce a surface charge at the boundary of the source. If a current validates these general conditions, all the exact magnetic moments and the ordinary ED become zero, and the source is completely characterized by time-odd, spatial-odd moments, i.e., toroidal moments. The fields radiated by a pure TD in the far zone do not differ from a conventional point ED. However, within the smallest spherical shell enclosing the source, the topology of the electromagnetic fields is drastically altered. Although we focus on the case of the dynamic TD, the same rules apply for dynamic toroidal quadrupoles, octupoles, and so forth.

Second, based on our enhanced physical insight into the problem, we design and verify experimentally for the first time a *pure*, *broadband,* dynamic TD source with all the other multipoles, including the ordinary ED, suppressed by more than three orders of magnitude. Unlike all previous works, our antenna displays a dynamic TD spanning hundreds of MHz in the microwave frequency range since we do not rely on any resonant behavior. These results are validated through a direct probing of the internal fields *inside the source region* as well as far-field measurements for a broad range of frequencies.

In the last part of our study, following a similar design strategy as for the pure dynamic TD source, we demonstrate analytically and prove numerically the excitation of a novel type of nonradiating source, a 'toroidal anapole', originating from the destructive interference of the dynamic TD with the first order MSR of the structure.

By rigorously justifying the split between ordinary electric and dynamic toroidal moments, we conclude that the latter can be independently controlled from the former and thus deserve to be considered as separate, meaningful physical entities. Furthermore, in the near future, toroidal anapoles could play a unique role in the rapidly growing field of anapole electrodynamics[41].

**Spherical multipole moments**

First, we formulate the mathematical setting in which we base our analysis. In what follows, we will always consider a piecewise continuous current distribution $\mathbf{J}(\mathbf{r},t)$ bounded in a finite space region, such as $\mathbf{J}(\mathbf{r},t)=0$ everywhere outside $\Omega$. The source is embedded in a homogeneous medium. We keep in mind that the Helmholtz theorem allows separating any field into its transverse (divergence-free, denoted by $\perp$) and longitudinal (curl-free $\parallel$) components $\mathbf{J}(\mathbf{r},t)=\mathbf{J}_{\parallel}(\mathbf{r},t)+\mathbf{J}_{\perp}(\mathbf{r},t)$. Without loss of generality, we consider the Fourier transform of $\mathbf{J}(\mathbf{r},t)$ to yield monochromatic components, denoted as $\mathbf{J}(\mathbf{r},\omega)$. Only $\mathbf{J}_{\perp}(\mathbf{r},\omega)$ contributes to the electromagnetic fields radiated outside $\Omega$ at frequency $\omega$. $\mathbf{J}_{\perp}(\mathbf{r},\omega)$ can be further expanded into plane wave components:

$$\mathbf{J}_{\perp}(\mathbf{r},\omega) = (2\pi)^{-3/2} \int_{-\infty}^{\infty} \mathbf{J}_k(\mathbf{k},\omega)\exp(i\mathbf{k}\cdot\mathbf{r})d^3\mathbf{k}. \qquad (1)$$

The contribution of $\mathbf{J}_{\perp}(\mathbf{r},\omega)$ to radiation is given only by those components of its plane wave expansion with wavevector $|\mathbf{k}|=\omega/c$. Therefore, all radiating $\mathbf{k}$ vectors define a spherical surface. It then becomes possible to expand the radiating $\mathbf{J}_k$ in terms of vector spherical harmonics that form an orthonormal basis for transverse vector functions in the unit sphere, as

$$\mathbf{J}_k(\hat{\mathbf{k}}) = \sum_{lm} a_{lm} \mathbf{Z}_{lm}(\hat{\mathbf{k}}) + b_{lm} \mathbf{X}_{lm}(\hat{\mathbf{k}}), \qquad (2)$$

where $l$ is the total angular momentum, referred hereon as the *multipole order*, and $m$ its projection to the z-axis. $\mathbf{X}_{lm}$ and $\mathbf{Z}_{lm}$ are provided in Eqs.(6) of Ref.[4]. The $a_{lm}, b_{lm}$ are the *spherical multipole moments* of electric and magnetic type, respectively. Each spherical multipole is a spherical tensor of order $l$, where $l=1$ corresponds to the *dipole*, $l=2$ to the *quadrupole*, and so forth. Straightforward formulas for $(a,b)_{lm}$ in terms of the original $\mathbf{J}(\mathbf{r},\omega)$ can be obtained by exploiting the orthogonality of the vector spherical harmonics[42]. They can be converted into Cartesian tensors through a simple basis transformation [42] to yield more familiar expressions. However, we emphasize that in this work, the term *spherical multipole* refers not to the basis of the tensor itself but to the fact that the coefficients describe the weights of each vector spherical harmonic in the expansion of the current, as given by Eq.(2). Here, we are interested in the spherical electric dipole $\mathbf{d}$ (spherical ED), defined (in Cartesian coordinates) as

$$\mathbf{d} = -\frac{1}{i\omega}\left\{\int d^3\mathbf{r}\, \mathbf{J}_\omega j_0(kr) + \frac{k^2}{2}\int d^3\mathbf{r}\left[3(\mathbf{r}\cdot\mathbf{J}_\omega)\mathbf{r} - r^2\mathbf{J}_\omega\right]\frac{j_2(kr)}{(kr)^2}\right\}. \qquad (3)$$

$j_i(kr)$ is the *i*th spherical Bessel function of the first kind, and $\mathbf{J}(\mathbf{r},\omega)$ has been rewritten as $\mathbf{J}_\omega$ for compactness. Similarly, the spherical magnetic dipole moment $\mathbf{m}$ (spherical MD) is given by

$$\mathbf{m} = \frac{3}{2}\int d^3\mathbf{r}\,(\mathbf{r}\times\mathbf{J}_\omega)\frac{j_1(kr)}{kr} \qquad (4)$$

The expressions are valid for a source of arbitrary extension and suffice to describe electromagnetic radiation everywhere outside $\Omega$.

**The physical significance of toroidal moments**

Maxwell's equations are invariant under both space inversion $\mathcal{P}$ and time reversal $\mathcal{T}$. The first operation changes the sign of all polar vectors (e.g., position or electric field) while leaving pseudovectors such as the magnetic field invariant. Time operator $\mathcal{T}$ reverses the flow of time. Thus, any moment describing a current distribution in Maxwell-Lorentz electrodynamics can be classified according to the operations in the space-time group [9]. In principle, up to four possible combinations exist: the moments can be either even or odd under $\mathcal{P}$ or $\mathcal{T}$. However, in the absence of magnetic charges, obtaining a parity even, time even tensor in a localized source is impossible. We are left with three choices.

At this point we regard **m** in Eq.(4). It is clear that $\mathcal{P}[\mathbf{m}] = \mathbf{m}$, since the spherical MD is defined as a pseudovector. Conversely, the $\mathcal{T}$ operation reverses the current flow so that $\mathcal{T}[\mathbf{m}] = -\mathbf{m}$. The spherical MD is, therefore a parity-even and time-odd tensor. On the other hand, the spherical ED in Eq.(3) is odd under parity inversion $\mathcal{P}[\mathbf{d}] = -\mathbf{d}$. Similarly, $\mathcal{T}[\mathbf{d}] = \mathbf{d}$. We are missing an additional tensor which should be odd under parity and time reversal. To recover this tensor, we resort to the long wavelength limit and expand $\mathbf{J}_\omega$ in a Taylor series with respect to the origin, yielding

$$\mathbf{J}_\omega \approx -i\omega \mathbf{p}\delta(\mathbf{r}) + \nabla \times [\mathbf{m}_0 \delta(\mathbf{r})] - \mathbf{T}\nabla\delta(\mathbf{r}) + ...,  \quad (5)$$

where $\delta(\mathbf{r})$ is a delta distribution centered at $\mathbf{r} = 0$. The first term in Eq.(5) is proportional to the ordinary ED moment $\mathbf{p}$, while the second term is the ordinary MD moment $\mathbf{m}_0$. Mathematically, they correspond to the first order terms in a Taylor series of **d** and **m** in Eq.(3) and Eq.(4), respectively. The ordinary ED moment has the textbook expression:

$$\mathbf{p} = -\frac{1}{i\omega}\int_\Omega \mathbf{J}_\omega d^3\mathbf{r}. \quad (6)$$

The third term in Eq.(5) is the dynamic TD moment, given by

$$\mathbf{T} = \frac{1}{10}\int \left[(\mathbf{r}\cdot\mathbf{J}_\omega)\mathbf{r} - 2r^2\mathbf{J}_\omega\right] d^3\mathbf{r}. \quad (7)$$

In what remains of this work we will omit the adjective 'dynamic' and refer to **T** simply as 'TD'. Crucially, while **p** and $\mathbf{m}_0$ retain the same behavior under $\mathcal{P}$ and $\mathcal{T}$ as **d** and **m**, this is no longer the case for **T**, which *changes its sign under both space and time inversions*. In the long wavelength approximation, the TD is thus the necessary moment that completes the set of allowed sign permutations under space and time inversion. Since the TD is odd under parity inversion, it appears after **p** in the Taylor series of **d**. However, it is evident from our discussion that the TD describes an elementary current with different symmetry properties in comparison with the ordinary ED moment. Such a distinction cannot be made directly with the spherical ED moment, which groups together the contributions of moments with the same parity, regardless of their behavior under time inversion. The physical meaning of the TD moment is clear: *it describes a current distribution that changes its sign under both space and time inversions*. The same argument can be used to justify the physical distinction between 'ordinary' and 'toroidal' terms of electric multipole coefficients of a higher order [16].

**Conditions for an ideal TD source**

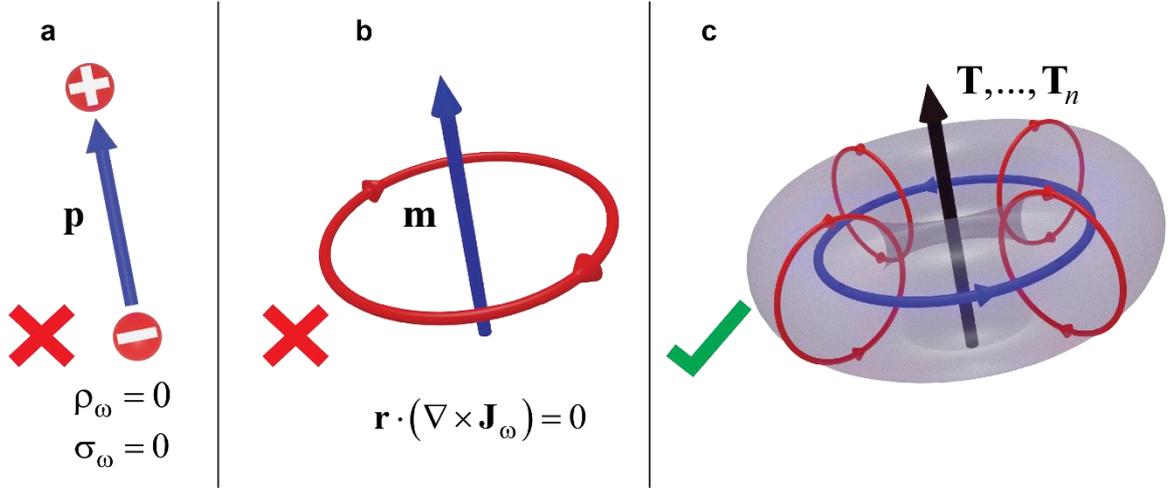

**Figure 1.** The conditions for any current distribution to be characterized entirely by toroidal moments. (a) Ordinary electric multipoles are suppressed in the absence of volume and *surface* charges. To the lowest order, the ordinary ED becomes precisely zero. (b) All magnetic moments, such as the MD depicted above, become zero if the curl of the current has no radial component (both ordinary and spherical). (c) If the conditions stated in (b) and (c) hold, the current is entirely described by toroidal moments and their MSR.

We derive here the conditions for a current distribution to have only toroidal character. We show that, in this special case, the ordinary contributions to the electric spherical multipoles vanish exactly, and the source can only support toroidal moments and their MSR. To do so, we impose the conditions separately for the moments to be simultaneously odd under space and time inversions.

We start with time inversion. Equivalently, this constraint can be understood as the *exact vanishing* of all ordinary electric moments. In the case of the TD, it is necessary to make $\mathbf{p}=0$, which has never been achieved in practice. To determine the conditions for this to happen, we rewrite the ordinary ED in Eq. (6) with the help of the continuity equation (refer to Supporting Information S1):

$$\mathbf{p} = -\frac{1}{i\omega}\int_{\Omega}\mathbf{J}_{\omega}d^3\mathbf{r} = \int_{\Omega}\mathbf{r}\rho_{\omega}d^3\mathbf{r} + \int_{\partial\Omega}\mathbf{r}\sigma_{\omega}d^2\mathbf{r} \ . \tag{8}$$

Eq. (8) shows that $\mathbf{p}$ can be expressed as a combination of a volume and a surface integral involving solely moments of the charge density. $\rho_{\omega}, \sigma_{\omega}$ are the Fourier transforms of the volume and surface charge densities, respectively. $\sigma_{\omega}$ arises at the boundary of the source, $\partial\Omega$. It is produced by any current having a discontinuity in its normal component at the boundary, $\mathbf{J}_{\omega}^{\perp}$. From the interface conditions for the current density (Supplementary Information S2), it follows that $\mathbf{J}_{\omega}^{\perp} = -i\omega\sigma_{\omega}$.

Eq. (8) explicitly shows the link between the ordinary multipoles associated with the bulk poloidal currents and those associated with the charge densities. This connection arises naturally from the condition of the source being finite and is not the consequence of any simplification. The latter expression is a manifestation of the well-known Siegert theorem in photonuclear physics, i.e., in the long wavelength limit, nuclear transitions are determined by the time rate of change of the electric dipole operator [44]. In the context of electrodynamics, it essentially tells us that the ordinary multipoles are proportional to the moments of the charge density. Since we are considering a pure current distribution in the absence of charges, $\rho_\omega = 0$, and the ordinary ED is simply $\mathbf{p} = \int_{\partial\Omega} \mathbf{r}\sigma_\omega d^2\mathbf{r}$.

We now arrive to our first important conclusion: a straightforward strategy to make $\mathbf{p}$ vanish consists of imposing $\sigma_\omega = 0$, or equivalently $\mathbf{J}_\omega^\perp = 0$, everywhere in $\partial\Omega$. With this constraint, the moments describing our current are forced to be odd under time-reversal (see Figure 1a).

Next, we impose the moments to be parity-odd. As follows from the discussion in the previous section, this translates into making all magnetic moments vanish, as depicted in Figure 1b. By inspecting the general form of a spherical magnetic moment of arbitrary order (Supplementary Information S5), we show that $\mathbf{J}_\omega$ must satisfy:

$$\mathcal{L} \cdot \mathbf{J}_\omega = 0. \tag{9}$$

In Eq.(9), $\mathcal{L} = -i\mathbf{r} \times \nabla$ is the orbital angular momentum operator. To interpret Eq.(9) geometrically, it is instructive to recast it as

$$\mathbf{r} \cdot (\nabla \times \mathbf{J}_\omega) = 0. \tag{10}$$

It turns out that the curl of the desired current must be tangential to an arbitrary sphere centered at the origin.

In summary, we have shown the most general conditions that a bounded current distribution must fulfill to become an ideal toroidal source for the first time to our knowledge. In order to be odd under time reversal, it can support neither volume $\rho_\omega = 0$ nor surface charges $\sigma_\omega = 0$ at the boundary. This is equivalent to imposing that the current is divergence-less and $\mathbf{J}_\omega^\perp = 0$ everywhere in $\partial\Omega$. To be odd under space (parity) inversion, the current must fulfill Eq.(10), i.e. its curl cannot have a radial component. We emphasize that these conditions aren't subject to any simplifications. In the next section, we will use them to derive a class of currents behaving as ideal toroidal sources.

**Ideal sources of toroidal moments**

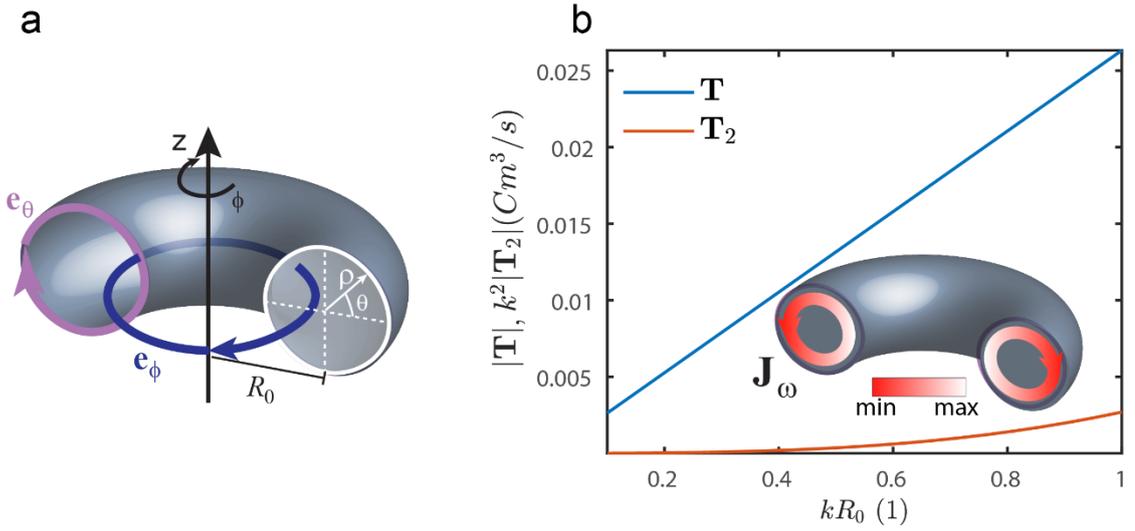

**Figure 2.** A current source with ideal toroidal character. (a) The toroidal/poloidal coordinate system used to obtain the exact analytical results in this section, particularly Eqs.(12)-(13). $\mathbf{e}_{\theta,\phi}$ represent the unit vectors in the poloidal (toroidal) directions. $\theta$ is the poloidal angle, while $\rho$ and $R_0$ are, respectively, the minor and major radius of a torus centered at the origin, and $\phi$ is the azimuthal angle. The unit vectors' orientation has been chosen to form a right-handed coordinate system. (b) Amplitudes of the TD and $1^{st}$ MSR ($\mathbf{T}_2$) for the current distribution in Eq.(11), with $K(\rho) = I_a S(\rho, \rho_0)$, as explained in the main text. The following parameters were used for its convenient representation: $I_a = 1A$, $k = 1m^{-1}$, $\rho_0 = 0.2m$. Inset: Scheme depicting an example of a current source with ideal toroidal character, with stronger current density in the internal walls of the torus.

We consider an arbitrary, bounded current expressed in a 'toroidal' system of coordinates, schematically shown in Figure 2a. A poloidal current takes the general form $\mathbf{J}_\omega = A_\theta \mathbf{e}_\theta$, where $A_\theta$ can be an arbitrary function of the coordinates (vanishing at some finite radius $\rho_0$) and $\mathbf{e}_\theta$ is a poloidal unit vector, depicted in Figure 2a and defined in Supplementary Information S3. Conveniently, this current distribution already fulfills $\mathbf{J}_\omega^\perp = 0$, and is therefore a good starting point. However, it can have volume charges. It can be shown (see Supplementary Information S4), that imposing $\rho_\omega = 0$ and Eq.(9), leads to:

$$\mathbf{J}_\omega(\rho, \theta) = \frac{K(\rho)}{R_0 + \rho \cos\theta} \mathbf{e}_\theta . \tag{11}$$

Eq.(11) describes a general family of poloidal currents in a torus with an ideal toroidal character. In the limit when the current is infinitesimally small, i.e. $K(\rho) \propto \delta(\rho - \rho_0)$ and $kR_0 \ll 1$, Eq.(11) resembles the one originally proposed in early works of Afanasiev [45] as a prototypical TD. Interestingly, the current density in Eq. (11) is not constant along the torus; it becomes more intense in the inner radius and decreases in the outer one, as depicted in the inset of Figure 2b. This condition is crucial to keep the current free of ordinary electric moments since otherwise, the current along every loop would be uncompensated, leading to charge accumulation. Critically, the family of currents proposed here can be of *arbitrary spatial extension*. Regardless of the size of the torus supporting it, a current density of the form given in Eq.(11) is always completely characterized *only by toroidal moments and their MSR*.

To demonstrate this, we investigate an analytically solvable case. We consider that the current is localized inside a torus with minor radius $\rho_0$ and major radius $R_0$, distributed homogeneously along the $\rho$ direction (inset of Figure 2b). Then, $K(\rho) = I_a S(\rho, \rho_0)$, where $I_a$ is a constant (in Amperes) and $S(\rho, \rho_0)$ is a step function that vanishes for $\rho > \rho_0$. Remarkably, the ordinary ED moment in Eq. (6) yields exactly $\mathbf{p} = 0$. Similarly, all the magnetic moments are precisely zero, since the current fulfills Eq.(10). The toroidal moments and their MSR remain nonzero. For example, the elementary TD [Eq.(7)] takes the form:

$$\mathbf{T} = \frac{I_a \pi^2}{3} \rho_0^3 R_0 \mathbf{e}_z. \tag{12}$$

The 1$^{st}$ MSR of the TD (or second order toroidal dipole) was derived in [6], and evaluates to:

$$\mathbf{T}_2 = \frac{I_a \pi^2}{150} \rho_0^3 R_0 \left(3\rho_0^2 + 5R_0^2\right) \mathbf{e}_z. \tag{13}$$

Eqs.(12)-(13) are plotted in Figure 2b as a function of $kR_0$. Interestingly, the ratio between $\mathbf{T}_2$ and $\mathbf{T}$ scales with $\propto 3\rho_0^2 + 5R_0^2$. Obviously, shorter wavelengths and/or much larger sizes are required in order to obtain significant higher order toroidal MSR response. However, what's important is that, regardless of the extension of the source, the current is always completely characterized by toroidal moments.

Finally, it is also worth noting that Eq.(11) does not correspond to the current flowing along a toroidal solenoid. The latter have for a long time been regarded as promising candidates for the observation of an ideal TD [8,13,46,47]. However, as we show in Supplementary Information S10, for any solenoidal coil with nonzero helicity bent into a torus, a nonzero MD will be excited, which can even be comparable in magnitude with the TD. Hence, toroidal solenoids appear unsuitable for the task and should not be used as examples of ideal TDs.

We emphasize that the conclusions in this section are fundamentally new since, until now, the dynamic TD and its successive MSR were only analytically shown to correspond to current distributions confined to a point in space [48]. Conversely, now we have generalized these results to currents of arbitrary spatial extension, which are no longer 'idealized' toy models.

**Design of an intrinsic TD source**

We are now ready to envision a realistic structure presenting an ideal toroidal character. The most straightforward strategy is to find a toroidal-like geometry that can support a current as in Eq.(11). Intuitively, a poloidal current can be viewed as the one produced by N magnetic dipoles of equal magnitude arranged in a head-to-tail fashion. To implement this idea, we model N subwavelength metallic loops arranged in a torus (inset of **Figure 2b**), with a radius $r$. Since $\frac{r}{\lambda} \ll 1$, where $\lambda$ is the input wavelength, every isolated loop supports a circulating current distribution radiating as an ordinary MD given by $\mathbf{m} = I\mathbf{A}$, where $I$ is the current in the loop and $\mathbf{A}$ is a vector normal to it, with magnitude equal to its cross-sectional area[49]. Critically, the current in every antenna is independently excited through an input port located at their base, as depicted in the inset of **Figure 2a**. This strategy allows us to fully control the *relative phase* of each loop, ensuring a head-to-tail distribution of the magnetic dipoles along the torus. Earlier works lacked this additional degree of freedom, resulting in the appearance of parasitic ordinary EDs and high order quadrupole moments.

First, we start with the simulation of one loop and calculate its cartesian multipole decomposition (**Figure 2a**). The result confirms the expected dominant MD response but also reveals a weak point ED stemming from the port contribution, approximately an order of magnitude smaller. In the torus, however, the weak ED contribution of each loop is canceled out by the one in the opposite side. We further validate these results by inspecting the current within the loop and the near-field distribution (**Figure 3a**). It can be clearly seen how the field around the antenna has its highest intensity close to the metallic surface and distributes evenly around the wire following the circular current, except at the top and bottom. This small inhomogeneity is attributed to the port, acting as a capacitance.

We then construct a torus arrangement with N=8 loops, which suffice to emulate a smooth poloidal current (**Figure 2b**). We calculate the total power radiated by the structure and its multipolar decomposition in both the spherical and cartesian representations. Due to the symmetry of the current distribution, we note that the structure does not support a magnetic response, as predicted in the previous section. The contribution of the spherical ED moment is sufficient to reconstruct the radiated power visually. However, it gives no insight on the topology of the current distribution supported by the source besides its evident electric nature.

Strikingly, the cartesian representation reveals that the TD is entirely responsible for the radiated power in the whole spectral range studied, with the ordinary ED suppressed everywhere by more than three orders of magnitude. To the best of our knowledge, this behavior has never been observed before. Our design constitutes the first *broadband TD source*. Importantly, in the range of $kr$ studied, the TD contribution coincides well with the spherical ED, indicating the absence of higher order MSR.

It is also interesting to visualize the effect of symmetry breaking in the current. For that purpose, we 'turn off' the port in one of the loops. Since the current no longer validates Eq.(10), it can manifest a magnetic response. This is confirmed in the multipole decomposition in **Figure 2c**: the source displays a strong MD response comparable to the TD, which contributes significantly to the total radiated power.

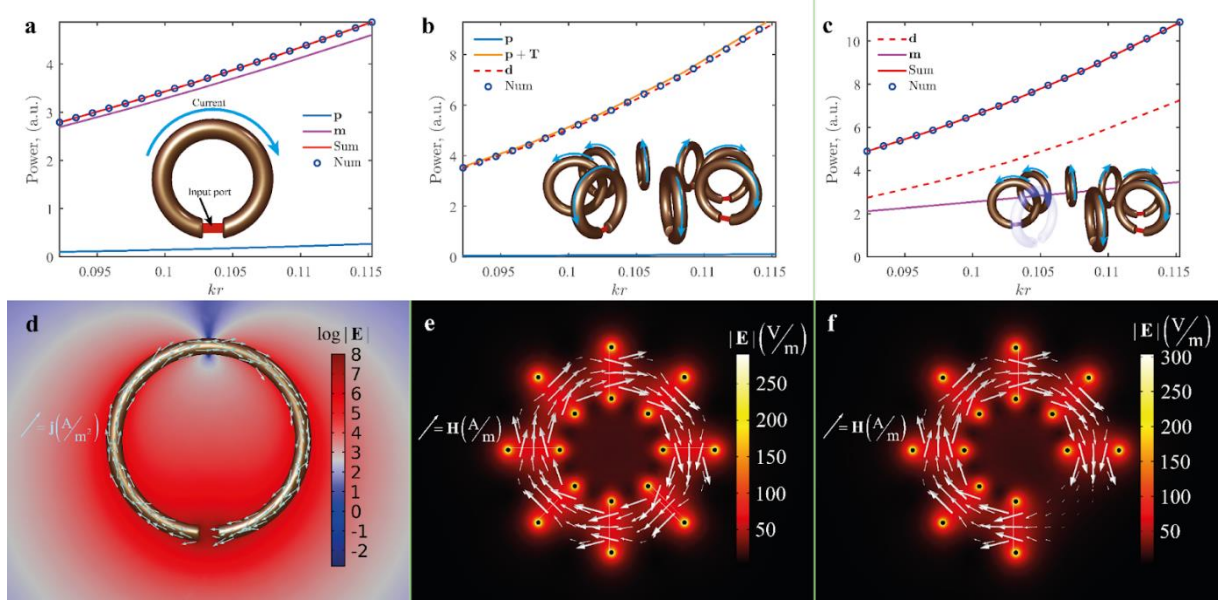

**Figure 3.** a) Illustration of one loop and multipole decomposition of radiated power with dominating magnetic dipole. B) System of 8 loops located in a way to form a torus with poloidal currents. The radiated power is almost completely described by a toroidal dipole with a small contribution from the toroidal MSR (the difference between orange and dashed red lines), while the ordinary ED is suppressed by more than three orders of magnitude in the whole range of $kr$ studied. C) The same system with one loop removed, resulting in broken symmetry and the appearance of a spherical MD comparable to the TD. (d) The loop used in simulations and the electric field (in logarithmic scale) it produces. Arrows show the current flowing on the surface of the loop. e) and f) show the electric field formed by loops with current (in normal scale) respectively for 8 loops and 7 loops with broken symmetry in the z=0 plane. The arrows show the magnetic field inside the 'discrete' torus formed by the loops.

To validate our results experimentally, we fabricated our structure and measured the near and far fields produced by it in the microwave frequency range (**Figure 4**) . The loops are made of 1 mm thick copper wire, which behaves as a PEC in the frequency range under consideration (inset of **Figure 4c**). In practice, the discrete ports feeding the toroidal source are implemented with a waveguide connected to a power divider, as shown in the photograph in the inset of **Figure 4c**. This strategy ensures each loop is supplied with the same amount of power and allows to control their relative phases. More details on the antenna and the experimental setup are given in Supplementary Information S8.

In **Figures 4a,c,** we provide a comparison of the simulated and measured magnetic fields for the frequency of 1000 MHz, where only the dominant azimuthal component is shown. The numerical and experimental results are in good agreement with each other. In both cases, we observe a circulating magnetic field inside the discrete torus formed by the loops, a clear signature of the TD. We emphasize that identical near-field patterns can be observed in a frequency range spanning hundreds of MHz, not only 1000 MHz. In the experiment, this has been tested up to 800 MHz (refer to Supplementary Information S9).

**Figure 4b** depicts the setup for the far field measurements. The antenna was installed on a numerically controlled positioning device, which allows rotating a sample mounted on its table with the precision of 0.1 degrees. The whole setup was placed in an anechoic chamber. A horn antenna with a lower frequency bound of 0.7 GHz was used as the far-field detector. The antenna was then placed 2.5 m away from the experimental model, which is enough to ensure the far-field region for the selected frequency range.

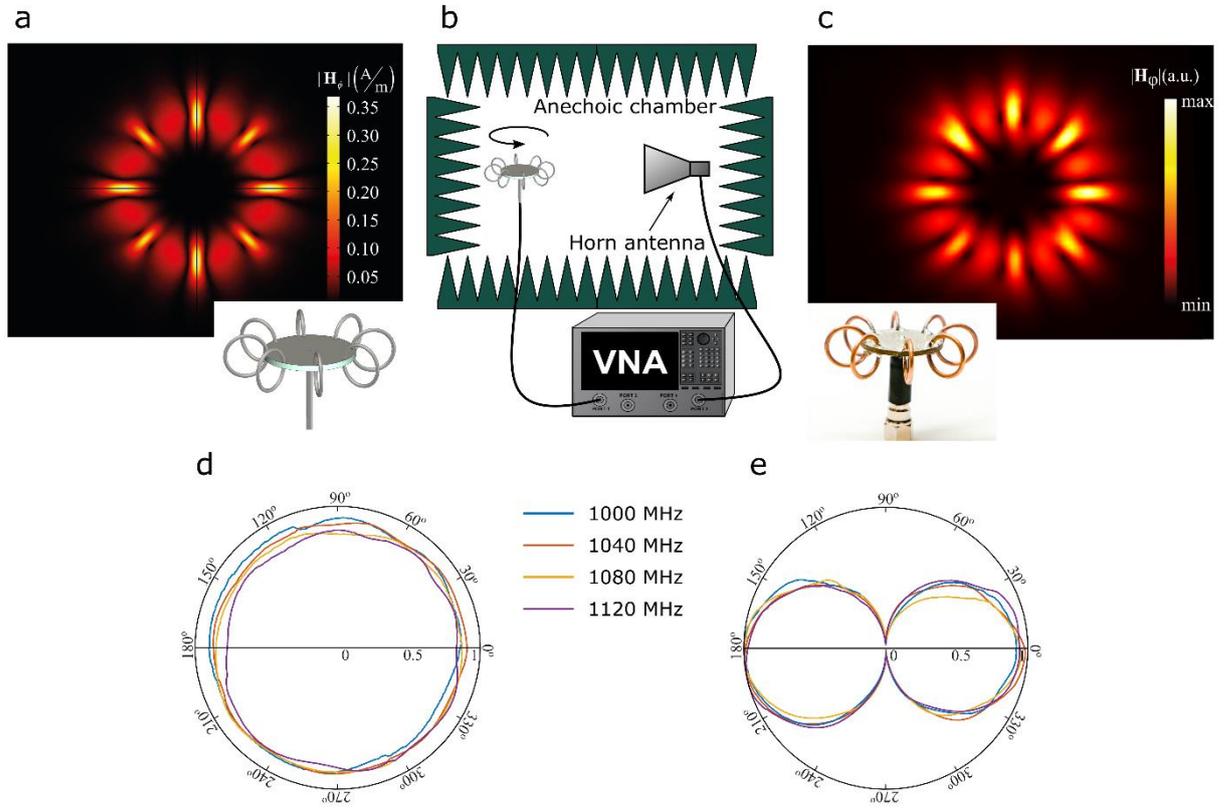

**Figure 4.** a) Numerical simulation of the azimuthal magnetic field distribution at a frequency of 1000 MHz in the plane perpendicular to the torus axis. Inset: a model of the antenna. B) Experimental setup for measuring the far-field. C). Measured azimuthal magnetic field distribution at frequency 1000 MHz. Inset: a picture of the manufactured experimental sample. D) Measured far-field directivity pattern in the plane of the torus, at four different frequencies, plotted in linear scale. E) Measured far-field directivity in the plane perpendicular to the torus, at four different frequencies, plotted in linear scale.

**Figures 4d-e** display the measured radiation patterns in and out of the torus plane, respectively. In all the frequencies studied, the source indeed displays a characteristic ED pattern associated with the TD excitation. This is further confirmed by an inspection of the near fields, being identical to the one in **Figure 4c** (not shown). The combination of near and far field measurements completely characterizes the source and confirms the first realization of a pure, broadband dynamic toroidal source.

Summarizing the results of this section, we have designed and implemented a realistic antenna behaving as an ideal toroidal source in the experiment. In stark contrast with previous works, which focused on resonant structures, we have been able to achieve *a broadband toroidal response* spanning hundreds of MHz in the microwave frequency range in the complete absence (three orders of magnitude suppression) of the ordinary ED moment, as well as all magnetic moments. This result has been confirmed semi-analytically with the help of both spherical and cartesian multipole representations, reaching a perfect agreement with the numerical calculations. Furthermore, we have directly measured both the near and far field signature of our pure toroidal source in good qualitative agreement with the simulations.

**Toroidal anapole**

The study of nonradiating charge-current configurations featuring nontrivial fields within a certain volume but zero at any other point in space is a fascinating research topic with a venerable history [50]. Nowadays, anapoles, the destructive interference of toroidal moments with their ordinary counterparts, have been extensively studied within the framework of nanophotonics [34,51,52].

The anapole takes place when the ordinary ED is cancelled by the TD according to $\mathbf{p} = -ik/c\mathbf{T}$, assuming that the MSR is negligible. Recently, however, the authors of Ref.[13] tentatively introduced the concept of a *toroidal nonradiating source*, i.e., where the destructive interference would not involve any ordinary multipole and would take place mainly between a toroidal moment and its MSR. Its realization could pave the way toward interesting developments in metamaterials and photonics exploiting the nontrivial effects of nonradiating sources, such as the Aharonov-Bohm effect [11]. Within this work, for brevity, we shall call the latter a toroidal anapole (TA), in contrast with the conventional ED anapole (EDA). To the best of our knowledge, the TA remains purely a proposal and has not been studied theoretically beyond a toy model.

For the emergence of a TA, the contributions to radiation from the TD and the 1$^{st}$ MSR of the whole structure must vanish, so that [13]

$$\mathbf{T} = -k^2 \mathbf{T}_2.$$  (14)

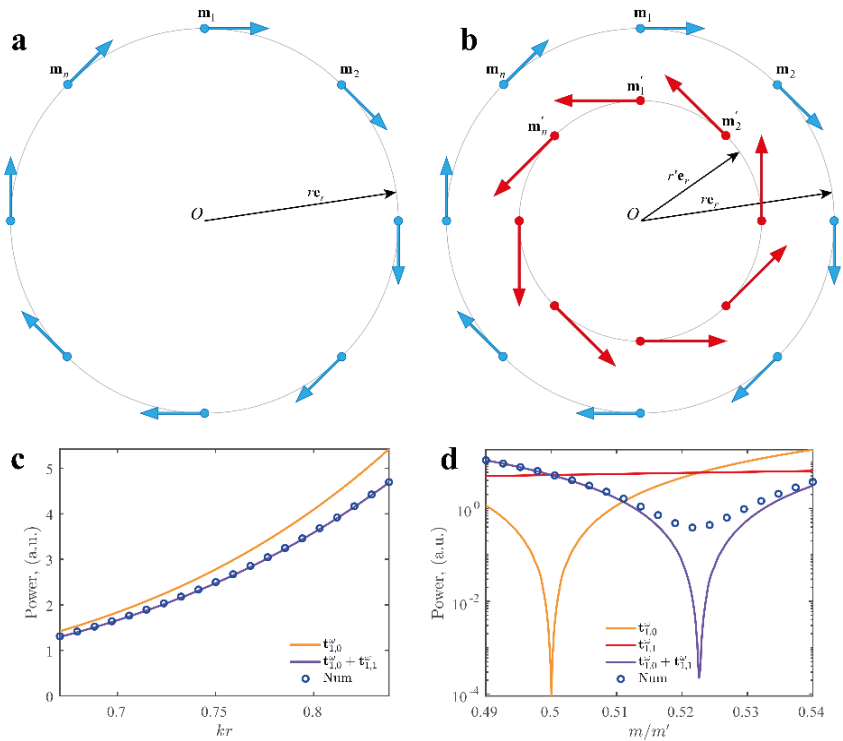

**Figure 5.** a) Schematic illustration of 8 PMDs. The PMDs are located equidistantly on the circle with a radius $r$. The magnitude of each dipole is $m$ while arrows show the direction of the magnetic moments.

b) Scheme of 16 PMDs arranged in 2 circular loops (8 PMDs on each circumference). Each circular array represents a 'discrete' version of the current in Eq.(11). The magnetic moments of the red PMDs are reversed in direction with respect to the blue ones, have magnitude $m'$, and are placed along a circumference with radius $r'$. c) Results of numerical simulation of dependence of power on $kr$ for 8 PMDs. Orange and purple lines are the power of TD and the sum of TD and 1$^{st}$ MSR, respectively. Blue circles correspond to the total power radiated by the two circular arrays. d) Radiated power vs. the ratio $m/m'$ for 16 PMDs. All legends are the same as in (c), and the red line is 1$^{st}$ MSR. Parameters: $kr = 0.24\pi$, r/r' = 2.

We consider a current allowing the 1$^{st}$ MSR to be comparable to the TD. From Figure 2b, one can appreciate that the TD and 1$^{st}$ MSR are quite different in magnitude for the current in Eq.(11). To enhance the MSR, we consider two concentric poloidal currents in two torii $a$ and $b$, with parameters $\rho_j, R_j, I_j$ where each of the currents satisfies Eq.(11). Eq.(14) can be solved analytically for our system, and leads to the relation (refer to the Supplementary Information S6):

$$I_a R_a \rho_a^2 \left(1 - \frac{k^2 P_a^2}{10}\right) = I_b R_b \rho_b^2 \left(1 - \frac{k^2 P_b^2}{10}\right). \tag{15}$$

In Eq.(15), $P_j^2 = R_j^2 + \rho_j^2$. $I_a$ and $I_b$ are the intensities of the currents circulating along the surface of each torus. To envision a system where the TA can be observed, we consider a circular arrangement of N point MDs (PMDs), depicted in Figure 5b.

Assuming an equal number of PMDs in both loops, in an analogous fashion to Eq.(15), the condition for the TA reduces to (Supplementary Information S7):

$$rm\left(1 - \frac{1}{10}k^2 r^2\right) - r'm'\left(1 - \frac{1}{10}k^2 r'^2\right) = 0, \tag{16}$$

where $r$, $r'$ are the radii of the largest and the smallest loop and $m$, $m'$ are the magnitudes of PMDs on the largest and the smallest loop. To validate Eq.(16), we numerically simulate this model for power versus $m/m'$ for N=16 PMDs. The results of the numerical experiments are shown in fig.X(d). One can see that the lines representing the power of TD (blue) and 1$^{st}$ MSR (red) are crossing at the theoretically predicted point, the power of the sum of TD and 1$^{st}$ MSR (orange line) is zero, and the total power (dashed line) has a minimum at that point. Interestingly, at $m/m' = 0.5$, the TD vanishes exactly, giving rise to a well-defined MSR response.

In this section, we theoretically described and confirmed through numerical simulations a new type of non-radiating source, the TA. Since our model of the TA is implemented with PMDs, clear candidates for its practical realization at the nanoscale are *rare-earth ions* which can be used as magnetic dipole emitters [53]. Well-known examples of such emitters are europium ions (Eu$^{3+}$) and erbium ions (Er$^{3+}$) [42]. Alternatively, the proposed structure could be implemented with an array of resonant metamolecules with artificial magnetic response, either dielectric or plasmonic in nature.

## Conclusion

Toroidal moments have been a subject of controversy ever since their introduction to electrodynamics. Until now, the separation of the spherical electric multipoles into ordinary and toroidal terms seemed somehow artificial and unnecessary for characterizing a current distribution. Here, we have shown that dynamic toroidal moments describe currents with a well-defined behavior under both space and time inversions. We have determined the exact conditions that any bounded current distribution must fulfill in order to be entirely characterized by dynamic toroidal moments and their MSR. Besides being divergence-less, we show that the curl of such currents must not have a radial component and importantly cannot induce a surface charge at the boundary of the source. If a current validates these general conditions, all the *exact* spherical magnetic moments and the ordinary ED become zero, and the source is completely characterized by time-odd, spatial-odd moments, i.e., toroidal moments and their MSR. The fields radiated by a pure TD in the far zone do not differ from a conventional point ED. However, within the smallest spherical shell enclosing the source, the topology of the electromagnetic fields is drastically altered. The distinction cannot, however, be made unless one has access to the electromagnetic field inside the source volume.

Secondly, based on our physical insight into the problem, we have designed and verified a pure, broadband dynamic TD source in the microwave frequency range experimentally for the first time. The toroidal response spans hundreds of MHz, in stark contrast with earlier attempts, where a pure TD could only be achieved at a single frequency [14,17]. Furthermore, all the other multipoles, including the conventional ED, are suppressed by more than three orders of magnitude.

In the last part of our study, following a similar design strategy as for the pure TD source, we have demonstrated analytically and proved numerically the excitation of a novel type of nonradiating source, a 'toroidal anapole', originating from the destructive interference of the TD with the first order MSR of the structure. Our results unambiguously prove the possibility of realizing an ideal toroidal source in electrodynamics. We expect our results to find application in several fields, from the design of novel nanophotonic devices to the characterization of biomolecules.

## Acknowledgments

The authors gratefully acknowledge the financial support from the Ministry of Science and Higher Education of the Russian Federation (Agreement No. № 075-15-2022-1150). The investigation of the novel toroidal anapole states has been partially supported by the Russian Science Foundation (Grant No. 21-12-00151). VB acknowledges the support of the Latvian Council of Science, project: DNSSN, No. lzp-2021/1-0048. LG acknowledges the financial support from the National Natural Science Foundation of China (92050104) and the Suzhou Prospective Application Research Project (SYG202039).

# Supplementary Information: On the existence of pure, broadband toroidal sources in electrodynamics

Adrià Canós Valero[1*], Dmitrii Borovkov[2], Mikhail Sidorenko[1], Pavel Dergachev[3,4], Egor Gurvitz[1], Lei Gao[5], Vjaceslavs Bobrovs[2], Andrey Miroshnichenko[6] and Alexander S. Shalin[2,5,7,8,9*]

[1]ITMO University, St. Petersburg 197101, Russia

[2]Riga Technical University, Institute of Telecommunications, Riga 1048, Latvia

[3]National Research University Moscow Power Engineering Institute 14 Krasnokazarmennaya st., Moscow 111250, Russia

[4]National University of Science and Technology (MISIS)
4 Leninsky pr., Moscow 119049, Russia

[5] School of Optical and Electronic Information, Suzhou City University, Suzhou 215104, China

[6] School of Engineering and Information Technology University of New South Wales Canberra Campbell, ACT 2600, Australia

[7]Faculty of Physics, M. V. Lomonosov Moscow State University, 119991 Moscow, Russia

[8]Center for Photonics and 2D Materials, Moscow Institute of Physics and Technology, Dolgoprudny 141700, Russia

[9]Kotelnikov Institute of Radio Engineering and Electronics, Ulyanovsk 432000, Russia

*adria.canos5@gmail.com

*alexandesh@gmail.com

## S1. Derivation of Eq.(8) of the main text

We will now show that, due to the source being finite, the first term in the right-hand side of Eq. is intimately connected with the longitudinal currents arising in the scatterer. For that purpose, we rewrite the expression of the electric dipole moment by making use of the identity [1]:

$$J_i = \nabla_j (r_i J_j) - r_i \nabla_j J_j \tag{S1}$$

the continuity equation

$$\nabla_j J_j = -\frac{\partial \rho_\omega}{\partial t}, \tag{S2}$$

and the interface condition for the current density, derived in section S2, which for a bounded source reads

$$\mathbf{J}_\omega(\mathbf{r}) \cdot \mathbf{n} = \mathbf{J}_\omega^\perp = -i\omega \sigma_\omega(\mathbf{r}), \tag{S3}$$

Where we denote as $\sigma_\omega$ the Fourier transform of the surface charge density. We can finally write the dipole coefficient as

$$\mathbf{p} = -\frac{1}{i\omega}\int d^3\mathbf{r}\, \mathbf{J}_\omega(\mathbf{r}) = \int \mathbf{r}\rho_\omega(\mathbf{r})d^3\mathbf{r} + \int \mathbf{r}\sigma_\omega(\mathbf{r})dS \tag{S4}$$

## S2. Interface conditions for the current density

The relation between the currents excited in two media joined by an interface can be simply derived analogously as the interface conditions for the electric and magnetic fields. We start with the conservation of charge in the integral form [2]:

$$\oint_{\partial \Omega} \mathbf{J} \cdot \mathbf{n}\, dS = -\frac{\partial}{\partial t}\oint_\Omega \rho\, d^3\mathbf{r} \tag{S5}$$

We now consider the bounding box depicted in Fig. S1, where the thickness $\delta$ is made arbitrarily small. Consequently, the volume charge density can be reduced to a surface charge density. The only contributions to the surface integral on the lhs correspond to the surfaces tangential to the boundary, yielding

$$\oint_{\partial \Omega} (\mathbf{J}_2 - \mathbf{J}_1) \cdot \mathbf{n}_{12}\, dS = -\frac{\partial}{\partial t}\oint_{\partial \Omega} \sigma\, dS \tag{S6}$$

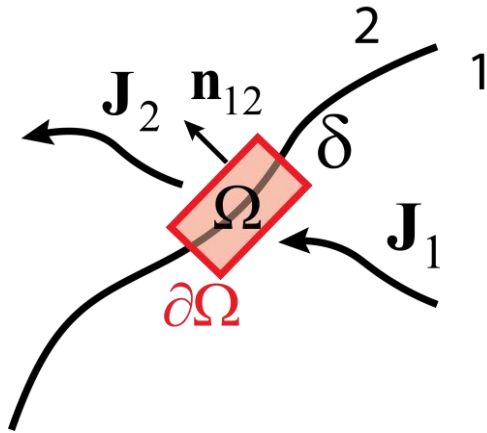

Figure S1. A small bounding box for the derivation of the interface conditions for the current density, traced around the interface between two media.

If the boundary is stationary, so that $\frac{D}{Dt}\partial\Omega = 0$, Eq. (S6) can be brought to the final expression:

$$(\mathbf{J}_2 - \mathbf{J}_1)\cdot\mathbf{n}_{12} = -\frac{\partial\sigma}{\partial t} \tag{S7}$$

If the current is bounded to medium 1 and the fields are time-harmonic, then Eq. (S7) establishes a straightforward link between the Fourier transforms of the surface charges and the internal current:

$$\mathbf{J}_\omega^1 \cdot \mathbf{n}_{12} = -i\omega\sigma_\omega. \tag{S8}$$

In the previous, the time dependence has been assumed of the form $e^{-i\omega t}$.

## S3. Toroidal Coordinates

By inspection of Fig.2a, one can obtain expressions for the Cartesian coordinates in terms of toroidal ones:

$$\begin{aligned} x &= (R_0 + \rho\cos\theta)\cos\phi \\ y &= -(R_0 + \rho\cos\theta)\sin\phi \\ z &= \rho\sin\theta \end{aligned} \tag{S9}$$

The evaluation of the integrals of the toroidal moments [e.g., Eq. (7) in the main text] can be simplified when they are performed in the toroidal coordinate system. For that purpose, we evaluate the Jacobian:

$$\left|J^{(x,y,z)\to(\rho,\theta,\phi,R_0)}\right| = -|J| = -\begin{vmatrix} \frac{\partial x}{\partial\rho} & \frac{\partial x}{\partial\theta} & \frac{\partial x}{\partial\phi} \\ \frac{\partial y}{\partial\rho} & \frac{\partial y}{\partial\theta} & \frac{\partial y}{\partial\phi} \\ \frac{\partial z}{\partial\rho} & \frac{\partial z}{\partial\theta} & \frac{\partial z}{\partial\phi} \end{vmatrix} = -\rho(R_0 + \rho\cos\theta) \tag{S10}$$

More explicitly $J$ is written as:

$$J = \begin{pmatrix} \cos\theta\cos\phi & -\rho\sin\theta\cos\phi & -(R_0 + \rho\cos\theta)\sin\phi \\ -\cos\theta\sin\phi & \rho\sin\theta\sin\phi & -(R_0 + \rho\cos\theta)\cos\phi \\ \sin\theta & \rho\cos\theta & 0 \end{pmatrix} \tag{S11}$$

The metrics of the toroidal coordinate system can be easily evaluated as [3]:

$$h_k = \left|\frac{\partial\mathbf{R}}{\partial q_k}\right| \tag{S12}$$

Where $q_1 = \rho$, $q_2 = \theta$, $q_3 = \phi$ so one needs to take the norm of each column in the Jacobian matrix, yielding:

$$h_\rho = 1$$
$$h_\theta = \rho \qquad\qquad\qquad\qquad (S13)$$
$$h_\phi = R_0 + \rho\cos\theta$$

The unit vectors are given by

$$\mathbf{e}_k = \frac{\partial \mathbf{R}}{\partial l_k} = \frac{\partial q_k}{\partial l_k}\frac{\partial \mathbf{R}}{\partial q_k} = \frac{1}{h_k}\frac{\partial \mathbf{R}}{\partial q_k} \qquad (S14)$$

Explicitly:

$$\mathbf{e}_\rho = \begin{pmatrix} \cos\theta\cos\phi \\ -\cos\theta\sin\phi \\ \sin\theta \end{pmatrix},\ \mathbf{e}_\theta = \begin{pmatrix} -\sin\theta\cos\phi \\ \sin\theta\sin\phi \\ \cos\theta \end{pmatrix},\ \mathbf{e}_\phi = \begin{pmatrix} -\sin\phi \\ -\cos\phi \\ 0 \end{pmatrix} \qquad (S15)$$

The scalar (inner) product of two vectors **u, v** in toroidal coordinates is

$$\mathbf{u}\cdot\mathbf{v} = u_\rho v_\rho + u_\theta v_\theta + u_\phi v_\phi \qquad (S16)$$

**Differential operators**

(i)     Divergence

The expression of the divergence in curvilinear coordinates is

$$\nabla\cdot\mathbf{A} = \frac{1}{h_1 h_2 h_3}\frac{\partial}{\partial q_i}\left(\frac{h_1 h_2 h_3}{h_i} A_i\right) \qquad (S17)$$

With Eqs. (S13) one may particularize it to

$$\nabla\cdot\mathbf{A} = \frac{1}{\rho(R_0+\rho\cos\theta)}\frac{\partial}{\partial q_i}\left(\frac{\rho(R_0+\rho\cos\theta)}{h_i} A_i\right) = \frac{1}{\rho f(\rho,\theta)}\frac{\partial}{\partial\rho}\left[\rho f(\rho,\theta) A_\rho\right] +$$
$$\frac{1}{\rho f(\rho,\theta)}\frac{\partial}{\partial\theta}\left[f(\rho,\theta) A_\theta\right] + \frac{1}{f(\rho,\theta)}\frac{\partial}{\partial\phi}(A_\phi) \qquad (S18)$$

With $f(\rho,\theta) = R_0 + \rho\cos\theta$. Simplifying the previous expression leads to

$$\nabla\cdot\mathbf{A} = \frac{\partial}{\partial\rho} A_\rho - \frac{1}{\rho(\rho\cos\theta + R_0)}\left(R_0 A_\rho + \rho A_\theta \sin\theta - \rho\frac{\partial}{\partial\phi} A_\phi\right) + \frac{2}{\rho} A_\rho + \frac{1}{\rho}\frac{\partial}{\partial\theta} A_\theta \qquad (S19)$$

(ii)     Gradient

The gradient in generalized coordinates is

$$\nabla A = \frac{\mathbf{e}_i}{h_i}\frac{\partial}{\partial q_i}(\mathbf{e}_j A_j) \qquad (S20)$$

The gradient operator can then be deduced to be

$$\nabla = \frac{\mathbf{e}_i}{h_i}\frac{\partial}{\partial q_i} \qquad (S21)$$

In toroidal coordinates, Eq. (S21) gives

$$\nabla = \frac{\mathbf{e}_i}{h_i}\frac{\partial}{\partial q_i} = \mathbf{e}_\rho \frac{\partial}{\partial \rho} + \frac{\mathbf{e}_\theta}{\rho}\frac{\partial}{\partial \theta} + \frac{\mathbf{e}_\phi}{R_0 + \rho\cos\theta}\frac{\partial}{\partial \phi} \quad (S22)$$

(iii)   Laplacian

Let $u(\rho,\theta,\phi)$ be an arbitrary, continuously differentiable scalar function. The Laplacian operator in toroidal coordinates can be derived as $\nabla \cdot \nabla u = \nabla^2 u$:

$$\begin{aligned}A_\rho &= \frac{\partial}{\partial \rho}u \\ A_\theta &= \frac{1}{\rho}\frac{\partial}{\partial \theta}u \\ A_\phi &= \frac{1}{R_0+\rho\cos\theta}\frac{\partial}{\partial \phi}u\end{aligned} \quad (S23)$$

$$\begin{aligned}\nabla^2 u &= \nabla \cdot \left(\mathbf{e}_\rho \frac{\partial}{\partial \rho}u + \frac{\mathbf{e}_\theta}{\rho}\frac{\partial}{\partial \theta}u + \frac{\mathbf{e}_\phi}{R_0+\rho\cos\theta}\frac{\partial}{\partial \phi}u\right) = \\ &\nabla \cdot \left(A_\rho \mathbf{e}_\rho + A_\theta \mathbf{e}_\theta + A_\phi \mathbf{e}_\phi\right) = \\ &\frac{\partial}{\partial \rho}A_\rho - \frac{1}{\rho(\rho\cos\theta+R_0)}\left(R_0 A_\rho + \rho A_\theta \sin\theta - \rho\frac{\partial}{\partial \phi}A_\phi\right) + \frac{2}{\rho}A_\rho + \frac{1}{\rho}\frac{\partial}{\partial \theta}A_\theta = \\ &\frac{\partial^2}{\partial^2 \rho}u - \frac{1}{\rho(\rho\cos\theta+R_0)}\left(R_0\frac{\partial}{\partial \rho}u + \sin\theta\frac{\partial u}{\partial \theta} - \frac{\rho}{R_0+\rho\cos\theta}\frac{\partial^2}{\partial^2 \phi}u\right) + \frac{2}{\rho}\frac{\partial}{\partial \rho}u + \frac{1}{\rho^2}\frac{\partial^2 u}{\partial^2 \theta}\end{aligned} \quad (S24)$$

The Laplacian is then

$$\nabla^2 = \left(\frac{\partial^2}{\partial^2 \rho} - \frac{R_0}{\rho(\rho\cos\theta+R_0)}\frac{\partial}{\partial \rho} + \frac{2}{\rho}\frac{\partial}{\partial \rho}\right) + \left(\frac{1}{\rho^2}\frac{\partial^2}{\partial^2 \theta} - \frac{\sin\theta}{\rho(\rho\cos\theta+R_0)}\frac{\partial}{\partial \theta}\right) + \frac{1}{(\rho\cos\theta+R_0)^2}\frac{\partial^2}{\partial^2 \phi} \quad (S25)$$

(iv)   OAM operator

To find the position vector $\mathbf{r}=(x,y,z)$, we first define the change of basis matrix between toroidal and cartesian coordinates:

$$A_{T\to C} = \begin{pmatrix} \cos\theta\cos\phi & -\sin\theta\cos\phi & -\sin\phi \\ -\cos\theta\sin\phi & \sin\theta\sin\phi & -\cos\phi \\ \sin\theta & \cos\theta & 0 \end{pmatrix} \quad (S26)$$

Where each column is simply each one of the unit vectors in (S15). Now the position vector is

$$\mathbf{r}_T = A_{C\to T}\begin{pmatrix}x\\y\\z\end{pmatrix} = \begin{pmatrix}\rho+R_0\cos\theta \\ -R_0\sin\theta \\ 0\end{pmatrix} = a_\rho \mathbf{e}_\rho + a_\theta \mathbf{e}_\theta \quad (S27)$$

With $A_{C\to T} = (A_{T\to C})^{-1}$. Taking into account the previous results, we can write explicitly $\mathbf{r}\times\nabla$:

$$\begin{aligned}
\mathbf{r} \times \nabla &= \left(a_\rho \mathbf{e}_\rho + a_\theta \mathbf{e}_\theta\right) \times \left(\mathbf{e}_\rho \frac{\partial}{\partial \rho} + \frac{\mathbf{e}_\theta}{\rho}\frac{\partial}{\partial \theta} + \frac{\mathbf{e}_\phi}{R_0 + \rho\cos\theta}\frac{\partial}{\partial \phi}\right) = \\
&\mathbf{e}_\rho \frac{a_\theta}{R_0 + \rho\cos\theta}\frac{\partial}{\partial \phi} - \mathbf{e}_\theta \frac{a_\rho}{R_0 + \rho\cos\theta}\frac{\partial}{\partial \phi} + \mathbf{e}_\phi\left(\frac{a_\rho}{\rho}\frac{\partial}{\partial \theta} - a_\theta \frac{\partial}{\partial \rho}\right) = \\
&-\mathbf{e}_\rho \frac{R_0 \sin\theta}{R_0 + \rho\cos\theta}\frac{\partial}{\partial \phi} - \mathbf{e}_\theta \frac{\rho + R_0\cos\theta}{R_0 + \rho\cos\theta}\frac{\partial}{\partial \phi} + \mathbf{e}_\phi\left(\frac{\rho + R_0\cos\theta}{\rho}\frac{\partial}{\partial \theta} + R_0 \sin\theta \frac{\partial}{\partial \rho}\right)
\end{aligned} \quad (S28)$$

Then $\mathcal{L} = -i\mathbf{r} \times \nabla$.

## S4. A divergence-less, poloidal current

We will now show how to derive the analytical current discussed in the main text. First, we impose the poloidal *geometry* of the current simply by assuming that it can be written as

$$\mathbf{J}_\omega = A_\theta \mathbf{e}_\theta, \tag{S29}$$

so that the current flows entirely along the meridians of an imaginary torus (we have omitted the arguments in $A_\theta$ for brevity). In the following, we will show that this is not a sufficient condition for the current to be considered "poloidal" in the differential sense.

Next, we calculate the divergence using Eq. (S19):

$$\nabla \cdot \mathbf{J}_\omega = -\frac{1}{\rho\cos\theta + R_0} A_\theta \sin\theta + \frac{1}{\rho}\frac{\partial}{\partial \theta} A_\theta \tag{S30}$$

The key step consists in setting the charge density in the source to zero, so that $\nabla \cdot \mathbf{J}_\omega = 0$ and

$$f(\theta) A_\theta + \frac{\partial}{\partial \theta} A_\theta = 0 \tag{S31}$$

with $f(\theta) = -\frac{\rho \sin\theta}{\rho\cos\theta + R_0}$. Eq. (S31) is a linear, homogeneous, first-order differential equation that can be easily solved by direct integration after the separation of variables:

$$\ln A_\theta = -\int f(\theta) d\theta, \tag{S32}$$

finally, we obtain

$$A_\theta = \frac{K(\rho,\phi)}{R_0 + \rho\cos\theta} \tag{S33}$$

This ends the proof for the first part of Eq. (12). We will now verify that the current is indeed poloidal by calculating $\mathcal{L} \cdot \mathbf{J}_\omega$. We combine Eqs. (S29), (S28) and (S33) obtaining (all terms disappear except the poloidal component):

$$\mathcal{L} \cdot \mathbf{J}_\omega = i \frac{\rho + R_0 \cos\theta}{(R_0 + \rho\cos\theta)^2}\frac{\partial}{\partial \phi} K(\rho,\phi) \tag{S34}$$

It is clear from Eq. (S34) that the necessary condition for $\mathcal{L} \cdot \mathbf{J}_\omega$ to become 0 is that $K(\rho,\phi) = K(\rho)$. Finally, after the previous discussion, the final form for the current distribution is:

$$\mathbf{J}_\omega = \frac{K(\rho)}{R_0 + \rho\cos\theta}\mathbf{e}_\theta \tag{S35}$$

Particularly, the function $K(\rho)$ can be a Heaviside function as in the main text, but we note that this is not necessary. The current will remain divergence-less and poloidal (in the differential sense) as long as it fits Eq.(S35).

To continue the analysis, it will be useful to know the formula for $\mathbf{r}\times\mathbf{J}_\omega$ in toroidal coordinates, which gives:

$$\mathbf{r}\times\mathbf{J}_\omega = \frac{1+\alpha\cos\theta}{\alpha+\cos\theta}K(\rho)\mathbf{e}_\phi \tag{S36}$$

with $\alpha = R_0/\rho$. The ordinary MD is proportional to the volume integral

$$\mathbf{m}_0 \propto \int_0^{2\pi}\mathbf{e}_\phi d\phi \int_{R_0}^R \rho K(\rho)\int_0^{2\pi}\left[\frac{1+\alpha\cos\theta}{\alpha+\cos\theta}\right](R_0+\rho\cos\theta)d\theta d\rho \tag{S37}$$

The integral $\int_0^{2\pi}\mathbf{e}_\phi d\phi = 0$ by symmetry, which proves that all basic magnetic dipoles cancel along the length of the torus. We note that this is the case only because $K$ is not a function of the azimuthal angle. Otherwise, there would be uncompensated MDs that would yield the overall coefficient nonzero.

However, the ordinary ED would still remain exactly zero since this depends only on the fact that the current is divergenceless. In the following section we will show that the condition $\mathcal{L}\cdot\mathbf{J}_\omega = 0$ imposed on the current indeed implies that *all magnetic moments of arbitrary order are exactly 0.*

### S5. Exact spherical magnetic moments

It is more convenient for our purposes to consider the exact form given in Jackson [2] (in the spherical basis). A magnetic moment of arbitrary order, $b_{lm}$, is given by the following expression:

$$b_{lm} = \frac{-k^2}{\sqrt{l(l+1)}}\int d^3\mathbf{r} j_l(kr)Y^*_{lm}(\hat{\mathbf{r}})\mathcal{L}\cdot\mathbf{J}_\omega(\mathbf{r}) \tag{S38}$$

Since our current fulfills Eq. (10) or Eq.(11), it follows directly that all $b_{lm} = 0$.

### S6. Derivation of the TA condition for two tori.

To begin with, we obtain an analytical expression for the MSR by expanding the exact electric dipole in a Taylor series for a small argument of $kr$. To do this, will use the expansion of the Bessel functions $j_0(kr)$ and $j_2(kr)$, then combine them into one sum. As a result, we get the expansion of the exact electric dipole in the Taylor series:

$$\mathbf{d} = -\frac{6}{i\omega}\int d^3\mathbf{r}\sum_{m=0}^\infty \frac{(-1)^m(m+1)(kr)^{2m}}{(2m+3)!}\left[(m+1)\mathbf{J}_\omega - m(\mathbf{r}\cdot\mathbf{J}_\omega)\frac{\mathbf{r}}{r^2}\right]. \tag{S39}$$

For the current as a starting point, we choose a bit simplified current from Eq.(12), where the dependence on $\rho$ is described by $\delta$-function:

$$\mathbf{J}^{(TD)} = \frac{K\delta(\rho-\rho_0)}{R+\rho\cos\theta}\hat{e}_\theta. \tag{S40}$$

Next, we need to add a second term corresponding to an additional torus to the current. Expressing one torus in the coordinates of another is rather problematic and not intuitive. Therefore, for greater clarity, we will express it in cylindrical coordinates $r, \psi, z$:

$$\begin{aligned} r_c &= R + \rho\cos\theta, \\ \psi &= -\phi, \\ z &= \rho\sin\theta. \end{aligned} \tag{S41}$$

$\rho$, $\sin\theta$ and $\cos\theta$ in cylindrical coordinates:

$$\begin{aligned} \rho &= \sqrt{(r_c-R)^2 + z^2}, \\ \sin\theta &= \frac{z}{\sqrt{(r_c-R)^2+z^2}}, \\ \cos\theta &= \frac{r_c-R}{\sqrt{(r_c-R)^2+z^2}}. \end{aligned} \tag{S42}$$

Cartesian components of vector $\hat{e}_\theta$ [Eq.(S15)] expressed in cylindrical coordinates:

$$\hat{e}_\theta = \frac{1}{\sqrt{(r_c-R)^2+z^2}} \begin{pmatrix} -z\cos\psi \\ -z\sin\psi \\ r-R \end{pmatrix}. \tag{S43}$$

Now we can write a current which, under the conditions discussed below, will support TA:

$$\mathbf{J}_\omega = \frac{K_0\delta(\rho(r_c,z)-\rho_0)}{r_c\sqrt{(r_c-R_0)^2+z^2}} \begin{pmatrix} -z\cos\psi \\ -z\sin\psi \\ r_c-R_0 \end{pmatrix} - \frac{K_1\delta(\rho(r_c,z)-\rho_1)}{r_c\sqrt{(r_c-R_1)^2+z^2}} \begin{pmatrix} -z\cos\psi \\ -z\sin\psi \\ r_c-R_1 \end{pmatrix}, \tag{S44}$$

Expressions of $\mathbf{r}\cdot\mathbf{J}_\omega$ and $|\mathbf{r}|^2$ are:

$$\mathbf{r}\cdot\mathbf{J}_\omega = -z\sum_{j=0}^{1}(-1)^j \frac{K_j R_j \delta(\rho(r_c,z)-\rho_j)}{r_c\sqrt{(r_c-R_j)^2+z^2}}, \tag{S45}$$

$$|\mathbf{r}|^2 = r_c^2 + z^2. \tag{S46}$$

First note that $x$ and $y$ components of all ED-like moments [Eq.(S39)] for the current in Eq.(S44) are zero. This is due to $\sin$ or $\cos$-like $\psi$-dependence of these components. $z$ components on the other hand need more detailed study.

TD:

$$\frac{ik}{c}(\mathbf{T})_\alpha = \frac{1}{10}\frac{k^2}{i\omega}\int d^3\mathbf{r}\left(2|\mathbf{r}|^2 J_\alpha - (\mathbf{r}\cdot\mathbf{J})r_\alpha\right), \tag{S47}$$

$$\frac{ik}{c}(\mathbf{T})_z = \frac{1}{10}\frac{k^2}{i\omega}(A_{0z}-A_{1z}), \tag{S48}$$

where for $j=0,1$,

$$A_{jz} = \int d^3\mathbf{r}\left(2(r_c^2+z^2)(r_c-R_j)+z^2 R_j\right)\frac{K_j\delta(\rho(r_c,z)-\rho_j)}{r_c\sqrt{(r_c-R_j)^2+z^2}}. \tag{S49}$$

Remember that $r_c = r_c(x,y) = \sqrt{x^2+y^2}$.

At this step, we can go back to the toroidal coordinates separately for each torus and integrate $\delta$-function. Integrating $\delta$-function leaves us surface integral with local coordinates $u$ and $v$ so that:

$$\begin{aligned} x &= (R_j+\rho_j\cos u)\cos v, \\ y &= -(R_j+\rho_j\cos u)\sin v, \\ z &= \rho_j\sin u. \end{aligned} \tag{S50}$$

With these substitutions, the integral can be cast into the form:

$$A_{jz} = \iint_\Omega du\, dv\left(2\left((R_j+\rho_j\cos u)^2+\rho_j^2\sin^2 u\right)\rho_j\cos u + R_j\rho_j^2\sin^2 u\right) \\ \frac{K_j}{\rho_j(R_j+\rho_j\cos u)}\sqrt{EG-F^2}, \tag{S51}$$

where

$$\begin{aligned} E &= (x_u')^2+(y_u')^2+(z_u')^2 = \rho_j^2, \\ F &= x_u'x_v'+y_u'y_v'+z_u'z_v' = 0, \\ G &= (x_v')^2+(y_v')^2+(z_v')^2 = (R_j+\rho_j\cos u)^2. \end{aligned} \tag{S52}$$

And the final result for integral in Eq.(S51) is:

$$A_{jz} = 10\pi^2 K_j R_j \rho_j^2, \tag{S53}$$

which gives the following expression for the $z$ component of TD [Eq.(S48)]:

$$\frac{ik}{c}(\mathbf{T})_z = \frac{\pi^2 k^2}{i\omega}\left(K_0 R_0 \rho_0^2 - K_1 R_1 \rho_1^2\right). \tag{S54}$$

In a similar fashion, using the following formula:

$$\frac{ik^3}{c}(\mathbf{T}_2)_\alpha = -\frac{1}{280}\frac{k^4}{i\omega}\int d^3\mathbf{r}\,|\mathbf{r}|^2\left(3|\mathbf{r}|^2 J_\alpha - 2(\mathbf{r}\cdot\mathbf{J})r_\alpha\right), \tag{S55}$$

which is received from Eq.(S39), one obtains for $z$ the component of 1$^{\text{st}}$ MSR:

$$\frac{ik^3}{c}(\mathbf{T}_2)_z = -\frac{1}{10}\frac{\pi^2 k^4}{i\omega}\left(K_0 R_0 \rho_0^2(R_0^2+\rho_0^2) - K_1 R_1 \rho_1^2(R_1^2+\rho_1^2)\right). \tag{S56}$$

The toroidal anapole state we are interested in is obtained with the condition $(\mathbf{T})_z + k^2(\mathbf{T}_2)_z = 0$, which gives:

$$K_0 R_0 \rho_0^2 \left(1 - \frac{k^2}{10}(R_0^2 + \rho_0^2)\right) - K_1 R_1 \rho_1^2 \left(1 - \frac{k^2}{10}(R_1^2 + \rho_1^2)\right) = 0. \qquad (S57)$$

Now if we replace $P_a^2 = R_0^2 + \rho_0^2$, $P_b^2 = R_1^2 + \rho_1^2$, $R_{0,1} = R_{a,b}$ and $K_0 = I_a$, $K_1 = I_b$ as a result, we will get Eq.(16) of the main text.

## S7. Derivation of the TA condition for a set of PMDs.

Here we will provide analytical proof for the results shown in Figure 5 of the main text for PMDs. Current for each PMD is defined as curl of magnetization multiplied by $\delta$-function. So overall current of this system is:

$$\mathbf{J} = \sum_j \mathbf{J}_j = \sum_j \nabla \times (\mathbf{m}_j \delta(\mathbf{r} - \mathbf{r}_j)). \qquad (S58)$$

To calculate an arbitrary electric dipole-like moment for the PMD with number $j$, we will put this expression into the integral in Eq.(S39). At first, we will ignore the prefactor and will add it later.

$$(\mathbf{T}_n)_\alpha \sim \int d^3\mathbf{r}\, r^{2n-2} \left((n+1) r_\mu r_\mu \varepsilon_{\alpha\beta\gamma} \partial_\beta m_\gamma^j \delta(\mathbf{r}-\mathbf{r}_j) - n r_\mu r_\alpha \varepsilon_{\mu\beta\gamma} \partial_\beta m_\gamma^j \delta(\mathbf{r}-\mathbf{r}_j)\right) \qquad (S59)$$

Calculating this integral for TD ($n=1$), one obtains the following result $-5\left[\mathbf{r}_j \times \mathbf{m}_j\right]_\alpha$, which after adding the prefactor gives:

$$\frac{ik}{c}\mathbf{T} = -\frac{1}{2}\frac{k^2}{i\omega}\sum_j [\mathbf{r}_j \times \mathbf{m}_j]. \qquad (S60)$$

For the 1st MSR integral in Eq.(S59) gives $14 r_j^2 \left[\mathbf{r}_j \times \mathbf{m}_j\right]_\alpha$. Adding the corresponding prefactor, we get:

$$\frac{ik^3}{c}\mathbf{T}_2 = \frac{1}{20}\frac{k^4}{i\omega}\sum_j r_j^2 [\mathbf{r}_j \times \mathbf{m}_j]. \qquad (S61)$$

An arbitrary toroidal MSR for the PMDs has the following form:

$$\frac{ik^{2n-1}}{c}(\mathbf{T}_n)_\alpha = \frac{(-1)^n}{i\omega}\frac{6n(n+1)k^{2n}}{(2n+2)!}\sum_j r_j^{2n-2}\left[\mathbf{r}_j \times \mathbf{m}_j\right]_\alpha. \qquad (S62)$$

Anapole state can be achieved by adding with two circles of PMDs, directed in opposite directions (Figure 5b). In principle, the condition for anapole state can be written as $(\mathbf{T} + k^2\mathbf{T}_2) + (\mathbf{T}' + k^2\mathbf{T}_2') = 0$, where TD and 1st MSR of the smaller circle of PMDs are primed. For a simpler case, we suppose the equal number of PMDs in both circles and equal absolute values of PMDs along every circle. The resulting condition for the anapole state is

$$rm\left(1 - \frac{1}{10}k^2 r^2\right) - r'm'\left(1 - \frac{1}{10}k^2 r'^2\right) = 0, \qquad (S63)$$

Corresponding to Eq.(17) of the main text. $r, r'$ are radii of the bigger and smaller circles and $m, m'$ are magnitudes of PMDs on bigger and smaller circles, respectively. The anapole state condition can be rewritten as $(\mathbf{T}+\mathbf{T}')+k^2(\mathbf{T}_2+\mathbf{T}_2')=0$. Analyzing this equation one finds that total TD of this system can become zero ($\mathbf{T}+\mathbf{T}'=0$) when $rm=r'm'$. First, it gives an opportunity to see 1st MSR in the absence of ED, TD and MD. Second, one can see that the necessary condition of trivial zero, when every multipole of one circle is combined with the same multipole of other circles, is: $m=m'$; $r=r'$.

Eq.(17) gives the following solutions for $m/m'$ and for $k$:

$$\frac{m}{m'}=\frac{r'}{r}\frac{1-\frac{1}{10}k^2 r'^2}{1-\frac{1}{10}k^2 r^2};$$

$$k=\left(\frac{10(rm-r'm')}{r^3 m - r'^3 m'}\right)^{\frac{1}{2}}.$$

(S64)

## S8. Details on the experimental implementation

Two parallel plates were used as a coaxially fed power divider to ensure a constant phase excitation of the rings. A schematic view of the experimental model is presented on **Figure 4b** of the main text. The rings have the diameter of 5.5 mm and are made of 1-mm-diameter copper wire. The power divider has the form of a parallel plate capacitor on a 1.5 mm thickness FR-4 dielectric substrate with a permittivity of 4.3 and loss tangent 0.02. The power divider is fed at its center point with a 50-Ohm coaxial cable, where the cable's inner conductor is soldered to the upper conducting plate of the power divider, and the shield is soldered to the lower plate. This divider, despite its low efficiency due to high reflection into the feed cable because of the impedance mismatch, provides an equal phase feeding of each ring of the field source. In the numerical simulations, carried out in a commercial microwave simulation software CST Microwave Studio, the model was fed by a Waveguide Port, attached to the cross-section of the coaxial cable. In the experimental verification, the coaxial cable was directly connected to the port of Vector Network Analyzer (VNA) Rohde-Schwarz ZVB-41. The model was installed on a numerically controlled positioning device, which allows rotating a sample mounted on its table with the precision of 0.1 degrees. The whole setup was placed in an anechoic chamber. A horn antenna with a lower frequency bound of 0.7 GHz was used as the far-field detector. The antenna was placed 2.5 m away from the experimental model, which is enough to ensure the far-field region for the selected frequency. The horn antenna was also connected to the VNA via the coaxial cable. The positioning device was then used to rotate the experimental model, and the measured value of the transmission coefficient S21 between the model and the receiving horn antenna, considered as a function of an angle, is proportional to the far-field directivity of the experimental field source model. The measurements were done for E- and H-planes of the whole 3-dimensional diagram. The results of the measurements, plotted in linear scale, are presented in **Figure 4d-e** of the main text for four different frequencies. The results of the numerical simulation and experimental measurements show the far-field diagram very close to the directivity diagram of a toroidal dipole, which is in full agreement with the theoretical results. Near-field measurement of the magnetic field in the XY plane was also carried out. A high-precision positioning device was used to place a small magnetic field probe in points forming a square grid with the step of 2 mm, and the transmission coefficient between the field source and the magnetic field probe antenna was measured for each point in the grid. The transmission coefficient is proportional to the amplitude of the tangential component of the magnetic field in the field probe location. A high-sensitive probe manufactured by Langer EMV-Technik was used.

## S9. Near field measurements at different frequencies

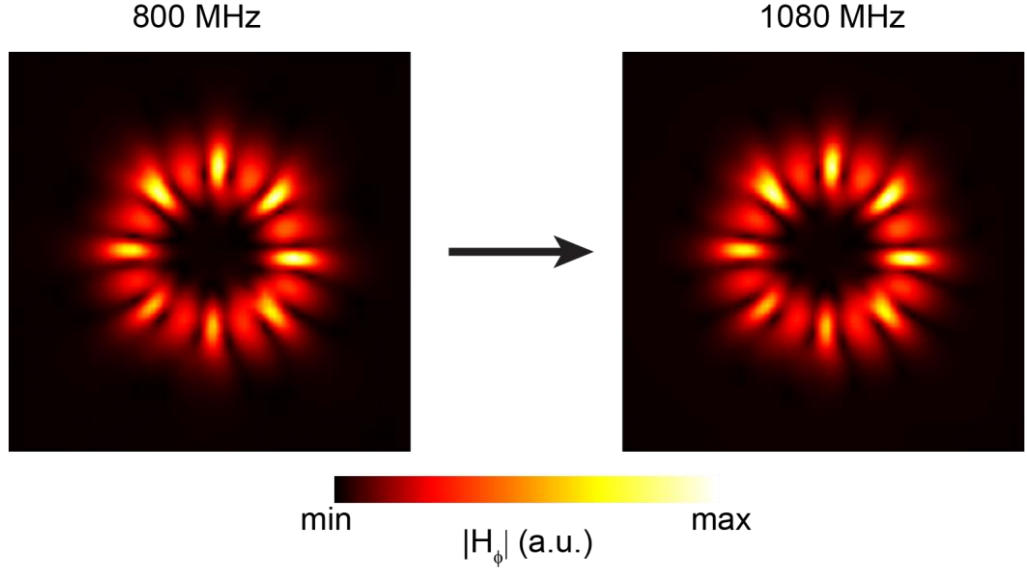

**Figure S2.** Near field measurements at different frequencies. Left: near field of the antenna at 800 MHz. Right: near field at 1080 MHz. The black arrow indicates that all near field measurements performed in the span of 280 MHz between the two plots are qualitatively identical to the two presented here. These results help confirm further that a strong TD is successfully excited in a very broad spectral range.

## S10. The multipole moments of toroidal solenoids

In this section, we argue that toroidal solenoids are not well suited for observing an ideal TD moment. First, consider an infinitely thin wire coil with N turns bent into a torus shape with a normal vector $\mathbf{n}$, external radius $R_0$ and internal radius $\rho_0$, along which circulates a homogeneous current $I$. In the toroidal coordinate system, such a current will follow a dependence of the form:

$$\theta(t) = \pi - Nt$$
$$\phi(t) = -t \quad , \tag{S65}$$

where $t$ is a free parameter. It can be shown that, if the current is homogeneous, it is indeed divergence-less and is therefore described only by time-odd multipoles (toroidal and magnetic). However, for any finite N, $\mathcal{L} \cdot \mathbf{J}_\omega \neq 0$, and hence magnetic moments are not zero. In particular, the ordinary MD evaluates to [4]:

$$\mathbf{m} = \pi R_0^2 I \left[ 1 + \frac{1}{2}\left(\frac{\rho_0}{R_0}\right)^2 \right] \mathbf{n},$$
(S66)

while the TD is given by

$$\mathbf{T} = \frac{\pi N I R_0 \rho_0^2}{2} \mathbf{n}.$$
(S67)

Only in the limit of vanishing helicity (tight bending), when $N \to \infty$, the TD becomes infinitely larger than the MD. Thus, in general, currents in toroidal solenoids are characterized by a mixture of toroidal and magnetic moments. Herein lies the reason, for instance, for the observation of chiral phenomena in metamaterials constituted of these structures [5]. In fact, electromagnetic chiral transitions cannot occur without magnetic response [6].

As an added note, there exist ways, in theory, to compensate for the additional magnetic moments by adding a wire loop or superposing two toroidal solenoids with opposite coil helicity [4]. Due to the complexity of the design, these ideas have never been implemented in practice, to the best of our knowledge. More importantly, the final structure is no longer a toroidal solenoid. In conclusion, a homogeneous current circulating along a toroidal solenoid cannot be considered a proper model of a TD moment.